\documentclass[conference]{IEEEtran}
\usepackage{balance}
\usepackage{booktabs}
\usepackage{multirow}
\usepackage{adjustbox}
\usepackage{float}
\usepackage{graphicx}
\usepackage[caption=true]{caption}
\usepackage[font=footnotesize]{subfig}
\usepackage{fixltx2e}
\usepackage{stfloats}
\usepackage{diagbox}
\usepackage{multirow}
\usepackage{colortbl}
\let\labelindent\relax
\usepackage{enumitem}
\usepackage{color}
\usepackage{multirow}
\usepackage{mathtools}
\usepackage{booktabs}
\newcommand{\squeezeup}{\vspace{-2.5mm}}
\newcommand{\squeezeupf}{\vspace{-1.5mm}}
 
\newcommand{\suiyi}[1]{\textcolor{black}{#1}}

\usepackage{cite}

\ifCLASSINFOpdf
\else
  \usepackage[dvips]{graphicx}
\fi
\begin{document}
%
\title{A Framework to Map VMAF with\\ the Probability of Just Noticeable Difference between Video Encoding Recipes}



%
\author{\IEEEauthorblockN{Jingwen Zhu\IEEEauthorrefmark{1}\textsuperscript{\textsection},
Suiyi Ling\IEEEauthorrefmark{1}\textsuperscript{\textsection}, 
Yoann Baveye\IEEEauthorrefmark{1},
Patrick Le Callet\IEEEauthorrefmark{1}}
\IEEEauthorblockA{\IEEEauthorrefmark{1}Nantes Université, Ecole Centrale Nantes, CNRS, LS2N, UMR 6004, F-44000 Nantes, France}
}
\maketitle

\begingroup\renewcommand\thefootnote{\textsection}
\footnotetext{Equal contribution}
\endgroup

\begin{abstract}
\suiyi{Just Noticeable Difference (JND) model developed based on Human Vision System (HVS) through subjective studies is valuable for many multimedia use cases. In the streaming industries, it is commonly applied to reach a good balance between compression efficiency and perceptual quality when selecting video encoding recipes. Nevertheless, recent state-of-the-art deep learning based JND prediction model relies on large-scale JND ground truth that is expensive and time consuming to collect. Most of the existing JND datasets contain limited number of contents and} are limited to a certain codec (\textsl{e.g.}, H264). \suiyi{As a result, JND prediction models that were trained on such datasets are normally not agnostic to the codecs. To this end, in order to decouple encoding recipes and JND estimation, we propose a novel  framework to map the difference of objective Video Quality Assessment (VQA) scores, \textsl{i.e., } VMAF, between two given videos encoded with different encoding recipes from the same content to the probability of having just noticeable difference between them. The proposed probability mapping model learns from} DCR \suiyi{test data, which is significantly cheaper compared to standard JND subjective test. As we utilize} objective VQA metric (\suiyi{e.g.,} VMAF that trained with contents encoded with different codecs) as proxy to \suiyi{estimate JND,  our model is agnostic to codecs and computationally efficient. Throughout extensive experiments, it is demonstrated that the proposed model is able to estimate JND values efficiently.}  
\end{abstract}


%
\IEEEpeerreviewmaketitle

\section{Introduction} 



\suiyi{It is well studied in the cognitive community through psychophysical experiments that our Human Visual System has a limited resolution. Just Noticeable Difference, defined as the minimum amount by which stimulus intensity must be adjusted in order to produce a noticeable difference, is commonly applied to represent this limitation. }
\suiyi{JND model developed based on HVS study has became the model of choice in many multimedia applications, especially for streaming industries, as it servers as a threshold for video compression platform. With the JND threshold, one can achieve high coding efficiency by eliminating perceptual redundancy without sacrificing the perceived quality to ensure best user experience.}  


\suiyi{\textbf{Subjective study:} To develop robust JND model, subjective studies need to be conducted to collect reliable JND data for the target usecase. Recently, several subjective studies have been conducted. Some JND datasets~\cite{jin2016statistical, liu2018jnd, fan2019picture, shen2020just} were released publicly for Picture Wise JND (PW-JND), and a few for Video Wise JND (VW-JND)~\cite{wang2016mcl, wang2017videoset}.} 
\suiyi{However, most of the existing subjective JND studies applied the `one-direction' JND search protocol, \textsl{i.e.}, by only decreasing/increasing the perceived quality of the compressed stimuli.}

\suiyi{\textbf{Objective JND models:} Since it is time consuming and costly to conduct subjective tests,}  especially for video contents, \suiyi{objective JND models that predict the JND values automatically are more desirable.} 
\suiyi{One of the most} common way\suiyi{s} to predict JND is to first predict \suiyi{the} Satisfied User Ratio (SUR) \cite{wang2018prediction, zhang2021deep, lin2020featnet, fan2019net}. For a given image/video content, \suiyi{the individual} JND \suiyi{of} different subjects \suiyi{are} different. SUR is the Complementary Cumulative Distribution Function (CCDF) of \suiyi{the} group-based JND annotations \suiyi{collected} from subjective test~\cite{wang2017videoset}. \suiyi{In the literature, t}he proxy of JND is usually \suiyi{the} encoding parameters (\textsl{e.g.}, QF, QP \suiyi{or the} CRF \suiyi{values} \suiyi{that decide the encoding setting}). 
Explicitly, given a certain encoding parameter $P$, SUR is \suiyi{defined as} the percentage of subjects who are satisfied with the \suiyi{the} compressed \suiyi{content}. \suiyi{In another word}, SUR is the \suiyi{ratio} of subjects that cannot perceive any difference between the reference/anchor and the \suiyi{content} encoded with $P$. JND can be derived from the SUR curve giving a certain threshold, \suiyi{and} the most commonly used JND threshold is $75\%$. \suiyi{Earlier, image JND/SUR models were developed for streaming contents. Liu \textsl{et al.}~\cite{liu2019deep} has proposed a deep learning based PW-JND model by formulating the task as a multi-category classification problem; Later, using also CNN, another PW-JND algorithm was presented in~\cite{tian2020just}; To overcome the challenge of training a deep PW-JND model with limited data, an effective JND model was proposed by Shen~\cite{shen2020just} using patch-level structural visibility learning; In~\cite{fan2019net}, a Siamese network was utilized to predict the SUR curve, which was later improved in~\cite{lin2020featnet}. Apart from the JND models dedicated for images, there are also some models were developed for videos. Zhang \textsl{et al.} modeled the SUR curve via Gaussian Processes Regression~\cite{wang2018prediction, wang2018analysis}; A new discrete cosine transform-based energy-reduced JND was demonstrated in~\cite{ki2018learning}. Similar to image SUR model, a video SUR model was proposed by Zhang \textsl{et al.} in~\cite{zhang2020satisfied}; Using deep learning, Zhang~\textsl{et al.} developed the Video Wise Spatial SUR method (VW-SSUR) for predicting the SUR value for compressed video along with the Video Wise Spatial-Temporal SUR (VW-STSUR) to boost the prediction accuracy.} 
\suiyi{Nevertheless}, most of the \suiyi{aforementioned} SUR prediction methods~\suiyi{(or JND approaches based on SUR estimation) suffer from} 2 drawbacks: (1) \suiyi{the} subjective JND datasets is indispensable. Specifically, large-scale JND datasets are necessary for deep learning-based method for SUR prediction \cite{fan2019net, lin2020featnet, zhang2021deep}. (2) \suiyi{these} SUR/JND \suiyi{models are} not codec agnostic. \suiyi{For instance}, the SUR prediction model proposed by \textsl{Wang et al.} \cite{wang2018prediction} \suiyi{that predicts} the QP value based on H.264 encoding datasets \cite{wang2017videoset} \suiyi{cannot be} directly \suiyi{applied} for other codec.  \squeezeup  

To address these \suiyi{issues} in existing works, rather than using encoding \suiyi{parameter} as \suiyi{the} JND proxy\suiyi{, which leads} to codec dependence, we~\suiyi{propose} to \suiyi{exploit} objective quality metric,~\suiyi{(\textsl{e.g.}, VMAF \cite{li2016toward}) to estimate the JND}. \suiyi{Intuitively, if an objective quality metric correlates well with human's opinions and is able to distinguish non-ambiguous pair confidently, it could be utilized to tell to which point the difference of a pair of stimuli becomes noticeable. Therefore, given a content as anchor with a certain perceived quality, the goal is to estimate the JND by calculating the difference of the objective quality scores between the anchor and another candidate with worse/better quality. With ground truth JND data, we can learn to which extend by increasing/decreasing the objective quality scores, we will notice the difference. Then, we can use this learned objective scores' difference (\textsl{e.g.,} difference in VMAF) as the new proxy for predicting the JND.} The anchor is not limited to the Source sequence (SRC), it can also be a Processed Video Sequence (PVS) \suiyi{obtained} with different encoding recipes. \suiyi{By increasing the quality of the anchor to a certain point (\textsl{i.e.,} the $1_{st}$ JND obtained via increasing the perceived quality), observer may be able to notice the quality got improved.  Vice versa, we can also get the JND through by decreasing the quality. In practical application, given the anchor, the proposed model outputs the difference of objective quality score $\Delta obj$ indicating to which point the change of quality becomes noticeable. One can then estimate objective quality score of the increased/decreased JND point $obj_{inc-JND}$ or $obj_{dec-JND}$, by simply adding/subtracting the $\Delta obj $ to/from the objective quality of the anchor $obj_{anchor}$. With the predicted $obj_{inc-JND}$ or $obj_{dec-JND}$, one can then map/interpolate the optimized codec configuration that generates the PVS with an objective quality score that equals (or close) to $obj_{inc-JND}$ or $obj_{dec-JND}$.}    \squeezeup 

\suiyi{The contribution of this study is two-fold: (1) we collected a JND dataset with a novel two-direction (with both increased and decreased quality) JND search subjective protocol, where each stimulus is also annotated with subjective DCR rating; (2) we presented} a framework to map the difference/residual of \suiyi{an objective quality metric (\textsl{e.g.,} VMAF)} between the anchor and a PVS with another encoding recipe \suiyi{by estimating} the probability of \suiyi{the PVS being considered with noticeable quality difference compared to the anchor with a mapping function}. \suiyi{It has to be highlighted that the} proposed framework takes subjective DCR test annotations as input but not \suiyi{the} subjective JND \suiyi{data, as JND subjective test is more resources and time consuming}. \suiyi{Interestingly, it is showcased that the calculated difference of objective scores for JND estimation depends on the quality range, where the objective score of the anchor content $obj_{anchor}$ falls in.}

\section{Methodologies} 
\subsection{Subjective \suiyi{studies}} \squeezeup  
\subsubsection{Two-directions JND search}\label{sec:jnd}
\begin{figure}[!t]
    \centering
    \includegraphics[width=2.8in]{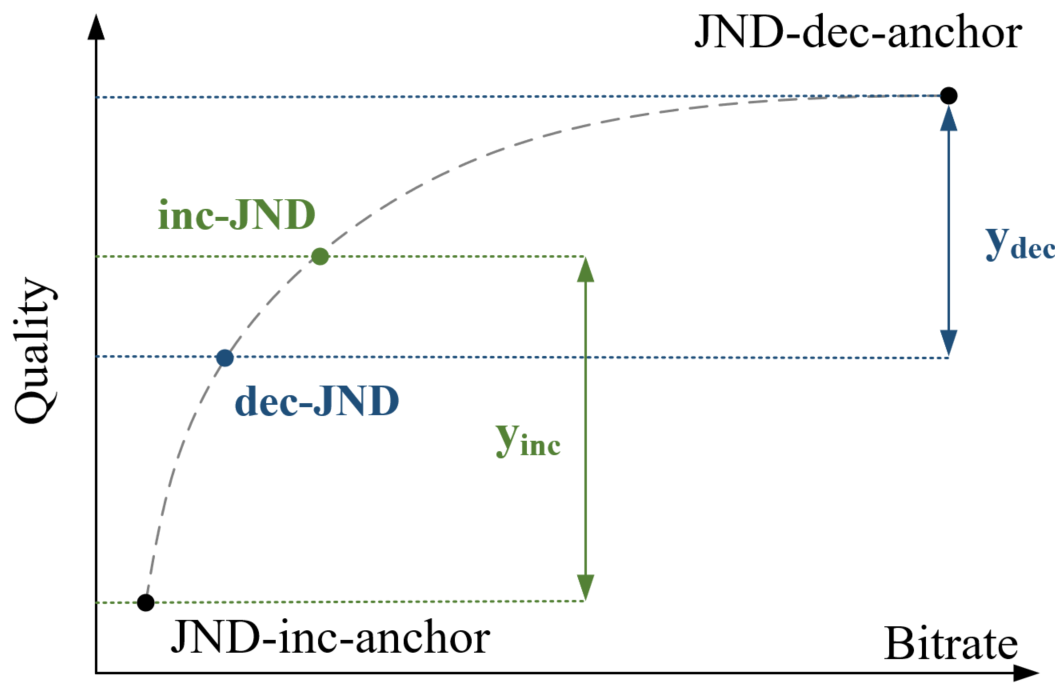}    \caption{Illustration of \suiyi{the proposed} 2-\suiyi{D} JND search}
    \label{fig_2_d} \squeezeup  \squeezeup \squeezeup 
\end{figure}

The objective of the two-direction (2-d) JND search is to \suiyi{provide the} JND \suiyi{ground truth} for validating the proposed \suiyi{JND} mapping function. 
\suiyi{As shown in Fig.\ref{fig_2_d}, the proposed 2-d JND search not only the JND with worse quality (\textsl{dec-JND}) compared to the anchor (\textsl{JND-dec-anchor}) via decreasing the perceived quality, but also the JND with better quality (\textsl{inc-JND}) compared to an anchor (\textsl{JND-inc-anchor}) by increasing the perceived quality. In our study, given a content, \textsl{JND-dec-anchor} is defined as the PVS/SRC with best quality, and \textsl{JND-inc-anchor} is the PVS that with the worse perceived quality (points from the same Rate-Quality (R-Q) curve). For each direction JND search, the bisection search process was applied, readers could refer to~\cite{wang2017videoset} for more details. By considering both directions, we can cover more streaming usecases, where \textsl{JND-dec-anchor} and/or \textsl{JND-inc-anchor} are/is required.}
 
20 HD SRCs and 10 UHD SRCs were selected from 229 \suiyi{pristine} videos from Amazon Prime Video \suiyi{streaming platform} based on the content selection ~\suiyi{strategy proposed} in \cite{ling2020towards} such that~\suiyi{the} selected contents \suiyi{are with wide-range of different} characteristics and  \suiyi{ambiguities in terms of quality}.
Each SRC was encoded with 39 $(3 \times 13)$ recipes (3 encoding resolution and each resolution \suiyi{with} 13 levels of distortion) using high efficiency video coding (HEVC).
The subjective test of 2-d JND search was conducted in a controlled lab environment. A 55-inch calibrated \suiyi{`}UHD Grundig Finearts 55 FLX 9492 SL\suiyi{'} was employed as \suiyi{the} display screen. The viewing distance \suiyi{was} set \suiyi{as 1.5$H$ for UHD and 3$H$ for HD contents} as recommended in ITU-R BT.2022~\cite{bt2012general}, where $H$ is the height of the screened video. 
\suiyi{Five} experts, namely \suiyi{the `}golden eyes\suiyi{',} participated in the 2-d JND search. They are familiar with encoding artifacts and \suiyi{thus the collected JND data is more accurate. Furthermore, less resources (time and observers) were required.}
\suiyi{ The} 2-d JND search\suiyi{es were} conducted \suiyi{for} each resolution respectively, (\suiyi{including} 1080p, 720p and 540p for HD). It \suiyi{is worth noting} that \suiyi{only} the \textsl{JND-dec-anchors} for 1080p of HD contents is the SRC, \suiyi{and} the \textsl{JND-dec-anchors} for 720p and 540p of HD contents \suiyi{are the PVS with best perceived quality. Similar setting was also applied for the UHD set}. 
For each resolution, the \suiyi{two anchors (\textsl{i.e.,} both the anchor for increased and decreased quality)} and their corresponding JNDs \suiyi{were} selected, \suiyi{which ended up to total} 12 $(4 \times 3)$ videos \suiyi{that are} with significantly different qualities for each content (11 PVS + 1 SRC).
 \squeezeup
\subsubsection{DCR based on 2-d JND search}\label{sec:dcr}

\suiyi{The proposed JND mapping model is based on estimating the probability that one stimulus is significant different/similar to another stimulus or not. Thus, subjective quality experiments are also required. To this end, we conducted a subjective quality study utilizing DCR protocol according to \textsl{ITU-T P.913} \cite{union2016methods} with the same contents selected in the 2-d JND searching test.} During the test, for \suiyi{each} observer, \suiyi{the SRC of each content was compared to 11 corresponding PVSs encoded with different encoding recipes. In total 24 observers participated in the test, and all of them} passed the pre-experiment vision check \cite{union2016methods} to ensure that they have correct visual acuity. 

 \squeezeup
\subsection{JND mapping framework} 
\begin{figure*}[htb]
    \centering
    \includegraphics[width=7in]{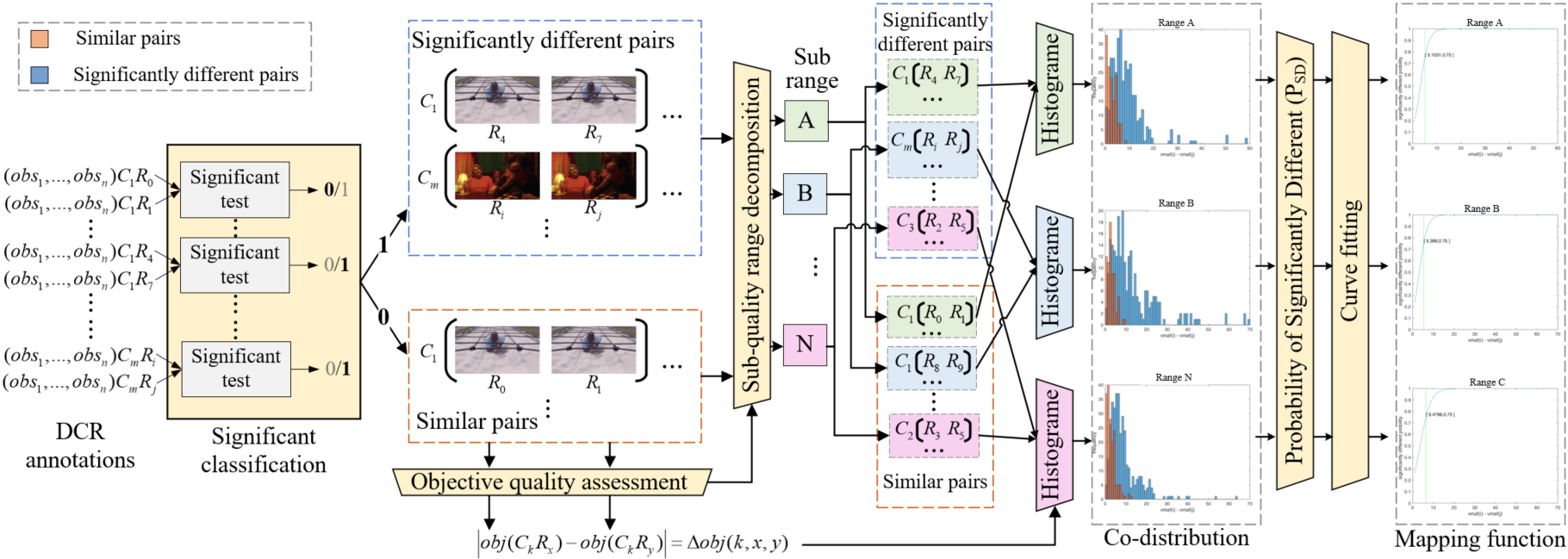}
    \caption{\suiyi{Diagram of the proposed framework of the JND mapping function using DCR datasets}} \squeezeup
    \label{fig:schema} \squeezeup  \squeezeup   \squeezeup
\end{figure*}
The JND prediction model is \suiyi{depicted} in Fig.\ref{fig:schema} \suiyi{(5 steps)}:
\suiyi{\textbf{Step 1 - Significant classification:}} For a given content ${C_k}$, \suiyi{$N$ JNDs (videos encoded from ${C_k}$ with different encoding recipes $R_x$)} were selected \suiyi{via the proposed} 2-\suiyi{D} JND search. \suiyi{From the $N$ selected videos (\textsl{i.e.}, increased/decreased JNDs), we formed pairs that took any two videos among the $N$ videos. In the end, there were in total $N$:$\frac{{N!}}{{d!(N - d)!}}$ pairs, where $d = 2$, $N = 12$. } 
For each content, all the formed pairs were classified into two classes: (1) significantly different pairs; (2) similar pairs. This classification \suiyi{was proceeded through a} significant test, \textsl{e.g.}, t-test, by using the individual subjective scores (opinion scores from all the observers $obs_1, ..., obs_n$) from the DCR subjective test described in Section \ref{sec:dcr}. Pairs with significant difference \suiyi{were} labeled with $sig =1$, otherwise $sig = 0$.

\suiyi{\textbf{Step 2 - Objective quality assessment:}}
The quality of each video \suiyi{was} evaluated by an certain objective quality assessment model $obj(\cdot)$ (\textsl{e.g.}, $obj=$ VMAF \cite{li2016toward}). Afterwards, the residuals/differences of objective quality scores for each pair \suiyi{were} calculated: $ \Delta obj(k,x,y) = \left| {obj({C_k}{R_x}) - obj({C_k}{R_y})} \right|$.

\suiyi{\textbf{Step 3 - Sub-quality range decomposition}} 
Before constructing the mapping function, video pairs ${C_k}\left\{ {{R_x},} \right.\left. {{R_y}} \right\}$ \suiyi{were grouped based on which sub-quality range the objective quality score of the anchor falls in. This grouping procedure that divides the entire quality range into a set of continuous sub-ranges is named as \textsl{sub-quality range decomposition}. The motivation behind} is that the perceptual distances between anchors and JND points differ along the quality range.
There are many possible ways to split the entire quality range. One of the most straightforward ways is to divide the whole quality range into equal units, 
 where the size of each bin is 5 in terms of VMAF score. Nevertheless, in this pilot test, such bins division strategy may end out into disconnected distribution (where some bins are empty) or distribution that contains only significantly different/similar pairs due to the limited numbers of tested videos. Therefore, in this study, the sub-ranges were split to have same amount of videos (balanced bins).   
%

\suiyi{\textbf{Step 4 - Mapping function design:}}
\suiyi{For each sub-quality range, the significantly different pairs that belong to this sub-range, were further represented by a histogram based on the $\Delta obj$ of the pair, \textsl{i.e.,} the objective score differences of the pairs}:
 $h_{dif}(\Delta obj)= \{ f_{1}^{dif}, \cdots , f_b^{dif}, \cdots,   f_{B}^{dif}  \}$ , where $f_b^{dif}$ is one bin that accumulates the frequency of pairs with objective quality residual equal to $b$, \textsl{i.e.,} $\Delta obj(k,x,y) = b$:\squeezeupf 
\begin{equation}
f_b^{dif} = \sum\limits_k {\sum\limits_{\{ x,y\} } {\mathbf{1}} } (\Delta obj(k,x,y) = b\;\& \;sig(k,x,y) = 1),  
\label{eq:freq_dif}
\end{equation}
where $\textbf{1}(c)$ is an indicator function that equals to 1 if the specified binary clause is true; and video pairs $\{ x,y\}$ belong to \suiyi{the current objective quality range}.  
Similarly, the distribution of similar pairs $h_{sim}(\Delta obj)$ could be obtained with:\squeezeupf 
\begin{equation}
f_b^{sim} = \sum\limits_k {\sum\limits_{\{ x,y\} } {\mathbf{1}} } (\Delta obj(k,x,y) = b\;\& \;sig(k,x,y) = 0).
\label{eq:freq_sim} 
\end{equation}
\suiyi{Afterwards, f}or each sub-quality range, we can get the two histograms, \suiyi{i.e.,} $h_{dif}(\Delta obj)$ and $h_{sim}(\Delta obj)$, with the same number and interval of bins of $\Delta obj$, namely, \suiyi{the} co-distribution. In this study, we utilized VMAF, \textsl{i.e.,} $obj$ = VMAF. As depicted in Fig.\ref{fig:schema}, the orange bar is the $h_{sim}(\Delta obj)$ and blue bar is the $h_{dif}(\Delta obj)$. It can be observed that similar pairs are distributed in regions where the $\Delta$VMAF is relatively small\suiyi{, and} significantly different pairs are distributed in regions with larger $\Delta$VMAF. For two \suiyi{different} encoding recipes $R_x$ and $R_y$ of \suiyi{the} content $C_k$, supposing that $\Delta \textrm{VMAF}(k,x,y) = b$, the probability that these two videos are Significantly Different (SD) \suiyi{is} calculated as: \squeezeupf 
\begin{equation}\label{eq:p_sd}
    {P_{SD}}(b) = \frac{{f_b^{dif}}}{{f_b^{dif} + f_b^{sim}}}, \squeezeupf 
\end{equation}
where $f_b^{dif}$ and $f_b^{sim}$ \suiyi{were} obtained from the co-distribution. 
\begin{figure}[!t]
    \centering
    \subfloat[Co-distribution]{\includegraphics[width=1.15in]{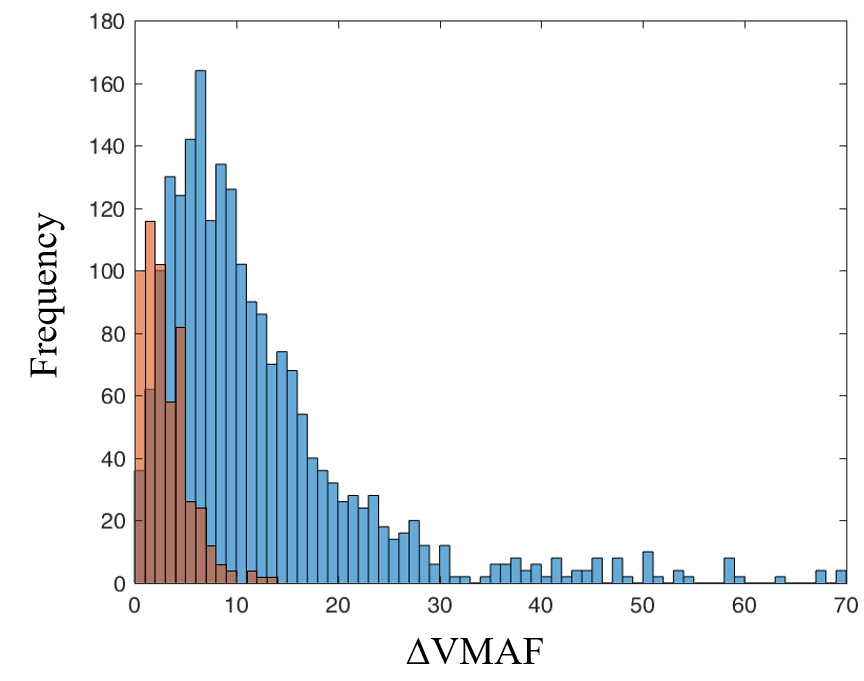}\label{sub_fig:co-distribution}}
    \subfloat[Original points of MF]{\includegraphics[width=1.15in]{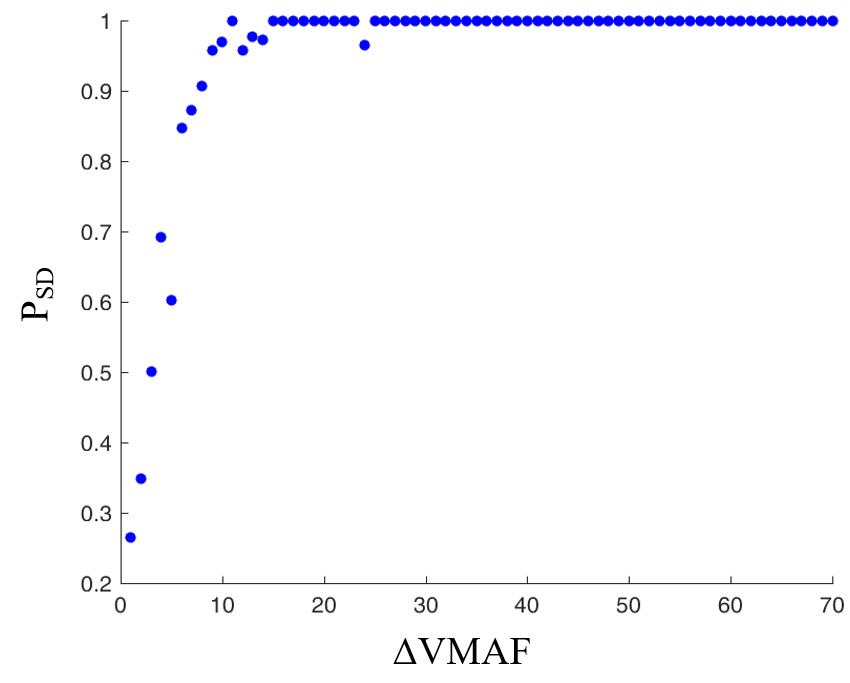}\label{sub_fig:original_points}}
    \subfloat[Fitted MF]{\includegraphics[width=1.15in]{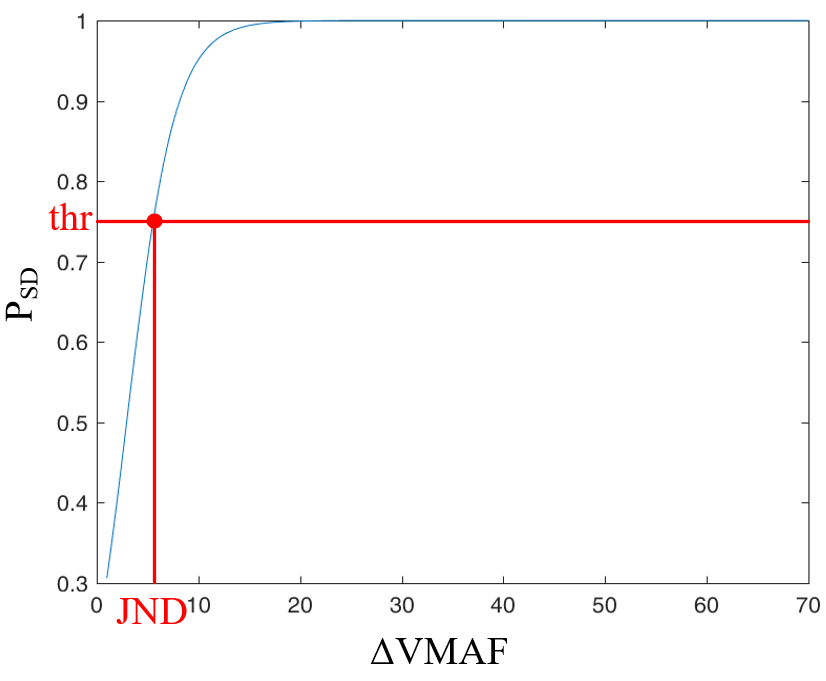}\label{sub_fig:fitted_mp}}
    \caption{How to obtain the mapping function (MF)}
    \label{fig:mapping_function} \squeezeup  
\end{figure}
Fig.\ref{fig:mapping_function} illustrate\suiyi{s} how to obtain the mapping function from the co-distribution: for each bin in the co-distribution (Fig.\ref{sub_fig:co-distribution}), \suiyi{the} value of $P_{SD}$ \suiyi{was obtained through} Eq.(\ref{eq:p_sd}). These $P_{SD}$ along with its corresponding $\Delta$VMAF are the original points of the Mapping Function (Fig.\ref{sub_fig:original_points}). The final Mapping Function \suiyi{was then} obtained by fitting these original points (Fig.\ref{sub_fig:fitted_mp}). 

\suiyi{\textbf{Step 5 - From mapping function to JND estimation:}}
After getting the mapping functions for different quality ranges, we can predict the JND of a given anchor \suiyi{with the corresponding estimated} $\Delta obj$ ( $obj$ = VMAF) by fixing a threshold. As illustrated in Fig.\ref{sub_fig:fitted_mp}, if the blue curve is the mapping function of \suiyi{an} anchor video $V_a$, and the threshold \suiyi{is set to} $thr = 75\%$, the corresponding value of $\Delta$VMAF $d$ is then \suiyi{considered} JND of $V_a$. In other words, for \suiyi{a PVS} $V_t$, \suiyi{which was encoded from the same SRC} as $V_a$, but encoded with \suiyi{a} different recipe, if $\left| {\textrm{VMAF}({V_a}) - \textrm{VMAF}({V_t})} \right| > b$, \suiyi{we assumed that observers} can perceive difference between \suiyi{them}. 

 \squeezeup 
\section{Experiment and Results}
\suiyi{The proposed JND estimation model was tested \suiyi{on both} the JND datasets collected in Section \ref{sec:jnd} and \suiyi{the} publicly available JND VideoSet \cite{wang2017videoset}. According to our best knowledge, we are the first to proposed such codec-independent JND estimation model using subjective quality data and objective quality metrics. Thus, as a preliminary study, we did not compared our model with any existing JND models.}
Before using the collected DCR annotations for the significant classification, 
three methods including (1) \textsl{VQEG HDTV Annex I} \cite{vqeg2009testplan}, (2) \textsl{ITU-R BT.1788} \cite{itu1999subjective} and (3) \textsl{ITU-R BT.500-12} \cite{bt2002methodology} were used to identify \suiyi{and remove} outliers\suiyi{(5 for HD and 3 for UHD)}. 
With significant t-test, we obtained 1075 significantly different pairs for HD and 563 significantly different pairs for UHD \suiyi{set}. We processed the collected HD and UHD datasets separately, \suiyi{due to page limitation, in this study, we }only \suiyi{present} the results of HD for demonstration. \suiyi{After unequal quality range decomposition (to ensure same number of pairs within each bin),} the \suiyi{decomposed} sub\suiyi{-}quality range\suiyi{s} for HD \suiyi{include} $(30,79]$, $(79, 86]$, $(86,90]$, $(90,95]$ and $(95,100]$. For video pair $\{ x,y\}$, it \suiyi{was} classified to \suiyi{the} sub\suiyi{-}quality range \suiyi{$A$,} if \suiyi{$\textrm{VMAF}(x) \in A \vee \textrm{VMAF}(y) \in A$.}

In our experiment\suiyi{s}, \suiyi{four} fitting functions \suiyi{were considered,} with the constraint that the function is monotonic on the full interval: (1) The 5-parameter logistic curve (5-para) \cite{tutorial2004objective}; (2) The 4-parameter cubic polynomial (4-para) \cite{tutorial2004objective}; (3) The 2-parameter logistic curve (2-para) \cite{tutorial2004objective}; (4) The Generalized Linear Model fitting (GLM) \suiyi{(commonly used in psycho-physical studies \cite{wichmann2001psychometric,lee2018generalized})}. 
To evaluate the performance of JND prediction, two commonly used regression error metrics, Mean Average Error (MAE) and Root Mean Square Error (RMSE) \suiyi{between the ground truth $y$ and the predicted label $\hat{y}$}, were \suiyi{calculated.}

The results on our JND dataset (\suiyi{Section \ref{sec:jnd}}) are summarized in Table \ref{tab:inc} for increasing JND and Table \ref{tab:dec} for decreasing JND. It can be observed that \suiyi{by applying the} GLM fitting function and \suiyi{a} threshold \suiyi{equals} to 0.95, \suiyi{our model achieves best performance in predicting both \textsl{dec-JND} and \textsl{inc-JND}.} 
To validate the generality of the proposed JND estimation model, the performance of JND prediction was also evaluated \suiyi{on the large-scale JND} VideoSet \cite{wang2017videoset}. It \suiyi{contains} 220 5-second SRCs in 4 resolutions. Each SRC \suiyi{was} encoded with H.264 codec with QP \suiyi{values} from 1 to 51. \suiyi{The dataset was labeled with the} 1st, 2nd and 3rd JND. The results are \suiyi{summarized} in Table \ref{tab:videoset}. It can be observed that the best performances are obtained when the threshold is 0.85 for \suiyi{all the three} JND\suiyi{s}. Furthermore, the prediction errors \suiyi{using the proposed model on the} VideoSet are slightly \suiyi{smaller} than \suiyi{ones on our 2-d JND dataset}, which \suiyi{further verifies the fact} that \suiyi{the} proposed framework is generic and codec agnostic.

\begin{table}[!t]
\centering
\caption{Performance of JND prediction ($\Delta$VMAF) for increasing JND for HD contents considering different thresholds $thr.$ and fitting functions $fit(\cdot)$}\label{tab:inc}
\arrayrulecolor{black}
\begin{adjustbox}{width=3in,center}
\begin{tabular}{!{\color{black}\vrule}c|cccc|} 
\hline
\multirow{2}{*}{\diagbox{$thr.$}{$fit(\cdot)$}}      & 5-para & 4-para & 2-para & GLM                               \\ 
\cline{2-5}
                                            & \multicolumn{4}{c|}{Mean Absolute Error (MAE)[1, 100]}       \\ 
\hline
0.75                                        & 5.1519 & 4.1997 & 5.1281 & 4.8915                            \\ 
\arrayrulecolor{black}
0.8                                         & 4.1956 & 3.9765 & 4.7420 & 4.4463                            \\ 

0.85                                        & 3.6163 & 3.8181 & 4.3771 & 3.9757                            \\ 

0.9                                         & 4.2271 & 3.7447 & 4.0420 & 3.6114                            \\ 

0.95                                        & 6.7110 & 3.8777 & 3.9551 & \textbf{3.4963}                   \\ 
\arrayrulecolor{black}\hline
\multicolumn{1}{!{\color{black}\vrule}c}{~} & \multicolumn{4}{c|}{Root Mean Square Error (RMSE) [1, 100]}  \\ 
\hline
0.75                                        & 7.2220 & 5.9346 & 7.0309 & 6.7227                            \\ 
\arrayrulecolor{black}
0.8                                         & 5.9573 & 5.7270 & 6.6600 & 6.2290                            \\ 

0.85                                        & 5.1240 & 5.5412 & 6.2459 & 5.7050                            \\ 

0.9                                         & 5.3592 & 5.3984 & 5.8011 & 5.1732                            \\ 

0.95                                        & 7.4985 & 5.4512 & 5.4493 & \textbf{4.8471}                   \\
\arrayrulecolor{black}\hline  
\end{tabular}
\end{adjustbox}
\end{table}
\squeezeup   

\begin{table}[!t] 
\centering
\caption{Performance of JND prediction ($\Delta$VMAF) for decreasing JND for HD contents considering different thresholds $thr.$ and fitting functions $fit(\cdot)$}\label{tab:dec}
\arrayrulecolor{black}
\begin{adjustbox}{width=3in,center}
\begin{tabular}{!{\color{black}\vrule}c|cccc|} 
\hline
\multirow{2}{*}{\diagbox{$thr$}{$fit(\cdot)$}}       & 5-para & 4-para          & 2-para & GLM                      \\ 
\cline{2-5}
                                            & \multicolumn{4}{c|}{Mean Absolute Error (MAE)[1, 100]}       \\ 
\hline
0.75                                        & 4.2586 & 3.4865          & 4.1996 & 4.0253                   \\ 
\arrayrulecolor{black}
0.8                                         & 3.6958 & 3.3277          & 3.8736 & 3.6564                   \\ 

0.85                                        & 3.8694 & \textbf{3.2493} & 3.6509 & 3.3475                   \\ 

0.9                                         & 5.1067 & 3.3602          & 3.6679 & 3.2665                   \\ 

0.95                                        & 7.6473 & 3.6988          & 4.2052 & 3.6863                   \\ 
\arrayrulecolor{black}\hline
\multicolumn{1}{!{\color{black}\vrule}c}{~} & \multicolumn{4}{c|}{Root Mean Square Error (RMSE) [1, 100]}  \\ 
\hline
0.75                                        & 7.5868 & 6.2914          & 7.2717 & 7.0435                   \\ 
\arrayrulecolor{black}
0.8                                         & 6.6783 & 6.1947          & 7.0217 & 6.6774                   \\ 

0.85                                        & 6.2428 & 6.1316          & 6.7698 & 6.3115                   \\ 

0.9                                         & 6.1184 & 6.1184          & 6.5555 & 5.9811                   \\ 

0.95                                        & 6.2585 & 6.2585          & 6.5525 & \textbf{5.8850}          \\
\arrayrulecolor{black}\hline   
\end{tabular}
\end{adjustbox} \squeezeup   
\end{table}

\begin{table}[!t]
\centering
\caption{Performance of JND prediction for the first, second, and third JND in \textbf{VideoSet} (1080p) considering different thresholds $thr.$ with GLM fitting functions.} \label{tab:videoset}
\begin{adjustbox}{width=\columnwidth,center}
\begin{tabular}{|c|cc|cc|cc|} 
\hline
$N\textsuperscript{th}$ & \multicolumn{2}{c|}{$1\textsuperscript{st}$ JND} & \multicolumn{2}{c|}{$2\textsuperscript{nd}$ JND} & \multicolumn{2}{c|}{$3\textsuperscript{rd}$ JND}  \\ 
\hline
$thr.$                  & MAE             & RMSE                         & MAE             & RMSE                         & MAE             & RMSE                          \\ 
\hline
0.75                  & 3.5026          & 4.6771                       & 3.5192          & 4.5879                       & 2.7160          & 3.6242                        \\
0.8                   & 3.1819          & 4.3186                       & 3.3584          & 4.3428                       & 2.5840          & 3.4218                        \\
0.85                  & \textbf{3.0491} & \textbf{4.1053}              & \textbf{3.3242} & \textbf{4.1961}              & \textbf{2.5691} & \textbf{3.3247}               \\
0.9                   & 3.2415          & 4.2470                       & 3.5338          & 4.2794                       & 2.7681          & 3.4638                        \\
0.95                  & 3.8400          & 4.7528                       & 3.5192          & 4.5879                       & 3.6060          & 4.5246                        \\
\hline
\end{tabular}
\end{adjustbox}
\end{table} \squeezeup   

\section{Conclusion}
In this work, \suiyi{a framework that maps the $\Delta$VMAF values to the probability of the JND between videos encoded with different encoding recipes is proposed. Given an anchor video as input, the proposed model takes into account which sub-quality-range the anchor video falls in, and output the estimated JND in terms of $\Delta$VMAF. With the $\Delta$VMAF, one can then select the right encoding recipes. } 
 \suiyi{The current JND prediction model} could be improved by taking the content characteristics into consideration, \suiyi{\textsl{e.g.,} using different mapping functions for different types of contents instead of based on quality range.} 



\balance
\bibliographystyle{IEEEtran}
\bibliography{refs}

\begin{thebibliography}{10}
\providecommand{\url}[1]{#1}
\csname url@samestyle\endcsname
\providecommand{\newblock}{\relax}
\providecommand{\bibinfo}[2]{#2}
\providecommand{\BIBentrySTDinterwordspacing}{\spaceskip=0pt\relax}
\providecommand{\BIBentryALTinterwordstretchfactor}{4}
\providecommand{\BIBentryALTinterwordspacing}{\spaceskip=\fontdimen2\font plus
\BIBentryALTinterwordstretchfactor\fontdimen3\font minus
  \fontdimen4\font\relax}
\providecommand{\BIBforeignlanguage}[2]{{%
\expandafter\ifx\csname l@#1\endcsname\relax
\typeout{** WARNING: IEEEtran.bst: No hyphenation pattern has been}%
\typeout{** loaded for the language `#1'. Using the pattern for}%
\typeout{** the default language instead.}%
\else
\language=\csname l@#1\endcsname
\fi
#2}}
\providecommand{\BIBdecl}{\relax}
\BIBdecl

\bibitem{jin2016statistical}
L.~Jin, J.~Y. Lin, S.~Hu, H.~Wang, P.~Wang, I.~Katsavounidis, A.~Aaron, and
  C.-C.~J. Kuo, ``Statistical study on perceived jpeg image quality via mcl-jci
  dataset construction and analysis,'' \emph{Electronic Imaging}, vol. 2016,
  no.~13, pp. 1--9, 2016.

\bibitem{liu2018jnd}
X.~Liu, Z.~Chen, X.~Wang, J.~Jiang, and S.~Kowng, ``Jnd-pano: Database for just
  noticeable difference of jpeg compressed panoramic images,'' in \emph{Pacific
  Rim Conference on Multimedia}.\hskip 1em plus 0.5em minus 0.4em\relax
  Springer, 2018, pp. 458--468.

\bibitem{fan2019picture}
C.~Fan, Y.~Zhang, H.~Zhang, R.~Hamzaoui, and Q.~Jiang, ``Picture-level just
  noticeable difference for symmetrically and asymmetrically compressed
  stereoscopic images: Subjective quality assessment study and datasets,''
  \emph{Journal of Visual Communication and Image Representation}, vol.~62, pp.
  140--151, 2019.

\bibitem{shen2020just}
X.~Shen, Z.~Ni, W.~Yang, X.~Zhang, S.~Wang, and S.~Kwong, ``Just noticeable
  distortion profile inference: A patch-level structural visibility learning
  approach,'' \emph{IEEE Transactions on Image Processing}, vol.~30, pp.
  26--38, 2020.

\bibitem{wang2016mcl}
H.~Wang, W.~Gan, S.~Hu, J.~Y. Lin, L.~Jin, L.~Song, P.~Wang, I.~Katsavounidis,
  A.~Aaron, and C.-C.~J. Kuo, ``Mcl-jcv: a jnd-based h. 264/avc video quality
  assessment dataset,'' in \emph{2016 IEEE International Conference on Image
  Processing (ICIP)}.\hskip 1em plus 0.5em minus 0.4em\relax IEEE, 2016, pp.
  1509--1513.

\bibitem{wang2017videoset}
H.~Wang, I.~Katsavounidis, J.~Zhou, J.~Park, S.~Lei, X.~Zhou, M.-O. Pun,
  X.~Jin, R.~Wang, X.~Wang \emph{et~al.}, ``Videoset: A large-scale compressed
  video quality dataset based on jnd measurement,'' \emph{Journal of Visual
  Communication and Image Representation}, vol.~46, pp. 292--302, 2017.

\bibitem{wang2018prediction}
H.~Wang, I.~Katsavounidis, Q.~Huang, X.~Zhou, and C.-C.~J. Kuo, ``Prediction of
  satisfied user ratio for compressed video,'' in \emph{2018 IEEE International
  Conference on Acoustics, Speech and Signal Processing (ICASSP)}.\hskip 1em
  plus 0.5em minus 0.4em\relax IEEE, 2018, pp. 6747--6751.

\bibitem{zhang2021deep}
Y.~Zhang, H.~Liu, Y.~Yang, X.~Fan, S.~Kwong, and C.~J. Kuo, ``Deep learning
  based just noticeable difference and perceptual quality prediction models for
  compressed video,'' \emph{IEEE Transactions on Circuits and Systems for Video
  Technology}, 2021.

\bibitem{lin2020featnet}
H.~Lin, V.~Hosu, C.~Fan, Y.~Zhang, Y.~Mu, R.~Hamzaoui, and D.~Saupe,
  ``Sur-featnet: Predicting the satisfied user ratio curve for image
  compression with deep feature learning,'' \emph{Quality and User Experience},
  vol.~5, no.~1, pp. 1--23, 2020.

\bibitem{fan2019net}
C.~Fan, H.~Lin, V.~Hosu, Y.~Zhang, Q.~Jiang, R.~Hamzaoui, and D.~Saupe,
  ``Sur-net: Predicting the satisfied user ratio curve for image compression
  with deep learning,'' in \emph{2019 eleventh international conference on
  quality of multimedia experience (QoMEX)}.\hskip 1em plus 0.5em minus
  0.4em\relax IEEE, 2019, pp. 1--6.

\bibitem{liu2019deep}
H.~Liu, Y.~Zhang, H.~Zhang, C.~Fan, S.~Kwong, C.-C.~J. Kuo, and X.~Fan, ``Deep
  learning-based picture-wise just noticeable distortion prediction model for
  image compression,'' \emph{IEEE Transactions on Image Processing}, vol.~29,
  pp. 641--656, 2019.

\bibitem{tian2020just}
T.~Tian, H.~Wang, L.~Zuo, C.-C.~J. Kuo, and S.~Kwong, ``Just noticeable
  difference level prediction for perceptual image compression,'' \emph{IEEE
  Transactions on Broadcasting}, vol.~66, no.~3, pp. 690--700, 2020.

\bibitem{wang2018analysis}
H.~Wang, X.~Zhang, C.~Yang, and C.-C.~J. Kuo, ``Analysis and prediction of
  jnd-based video quality model,'' in \emph{2018 Picture Coding Symposium
  (PCS)}.\hskip 1em plus 0.5em minus 0.4em\relax IEEE, 2018, pp. 278--282.

\bibitem{ki2018learning}
S.~Ki, S.-H. Bae, M.~Kim, and H.~Ko, ``Learning-based
  just-noticeable-quantization-distortion modeling for perceptual video
  coding,'' \emph{IEEE Transactions on Image Processing}, vol.~27, no.~7, pp.
  3178--3193, 2018.

\bibitem{zhang2020satisfied}
X.~Zhang, C.~Yang, H.~Wang, W.~Xu, and C.-C.~J. Kuo, ``Satisfied-user-ratio
  modeling for compressed video,'' \emph{IEEE Transactions on Image
  Processing}, vol.~29, pp. 3777--3789, 2020.

\bibitem{li2016toward}
Z.~Li, A.~Aaron, I.~Katsavounidis, A.~Moorthy, and M.~Manohara, ``Toward a
  practical perceptual video quality metric,'' \emph{The Netflix Tech Blog},
  vol.~6, no.~2, 2016.

\bibitem{ling2020towards}
S.~Ling, Y.~Baveye, D.~Nandakumar, S.~Sethuraman, and P.~Le~Callet, ``Towards
  better quality assessment of high-quality videos,'' in \emph{Proceedings of
  the 1st Workshop on Quality of Experience (QoE) in Visual Multimedia
  Applications}, 2020, pp. 3--9.

\bibitem{bt2012general}
I.~BT, ``General viewing conditions for subjective assessment of quality of
  sdtv and hdtv television pictures on flat panel displays,''
  \emph{International Telecommunication Union}, 2012.

\bibitem{union2016methods}
I.~Union, ``Methods for the subjective assessment of video quality audio
  quality and audiovisual quality of internet video and distribution quality
  television in any environment,'' \emph{SERIES P: TERMINALS AND SUBJECTIVE AND
  OBJECTIVE ASSESSMENT METHODS}, 2016.

\bibitem{vqeg2009testplan}
VQEG, ``Vqeg hdtv group : Test plan for evaluation of video quality models for
  use with high definition tv content (draft version 3.1),''
  \emph{\url{https://vqeg.github.io/software-tools/subjective\%20test\%20software/matlab-screening/}}.

\bibitem{itu1999subjective}
P.~ITU-T~RECOMMENDATION, ``Subjective video quality assessment methods for
  multimedia applications,'' 1999.

\bibitem{bt2002methodology}
R.~I.-R. BT, ``Methodology for the subjective assessment of the quality of
  television pictures,'' \emph{International Telecommunication Union}, 2002.

\bibitem{tutorial2004objective}
I.~Tutorial, ``Objective perceptual assessment of video quality: Full reference
  television,'' \emph{ITU-T Telecommunication Standardization Bureau}, 2004.

\bibitem{wichmann2001psychometric}
F.~A. Wichmann and N.~J. Hill, ``The psychometric function: I. fitting,
  sampling, and goodness of fit,'' \emph{Perception \& psychophysics}, vol.~63,
  no.~8, pp. 1293--1313, 2001.

\bibitem{lee2018generalized}
Y.~Lee, J.~A. Nelder, and Y.~Pawitan, \emph{Generalized linear models with
  random effects: unified analysis via H-likelihood}.\hskip 1em plus 0.5em
  minus 0.4em\relax Chapman and Hall/CRC, 2018.

\end{thebibliography}

\end{document}